\documentclass[twocolumn,showpacs,prb,preprintnumbers,superscriptaddress,amsmath,amssymb]{revtex4}
\usepackage{graphicx}
\usepackage{dcolumn}
\usepackage{bm}
\usepackage{amsmath,amssymb}
\newcommand{\be}{\begin{eqnarray}}
\newcommand{\ee}{\end{eqnarray}}

\begin{document}
\title{Almost mobility edges and existence of critical regions in one-dimensional quasiperiodic lattices}
\author{Yucheng Wang}
\affiliation{Beijing National
Laboratory for Condensed Matter Physics, Institute of Physics,
Chinese Academy of Sciences, Beijing 100190, China}
\author{Gao Xianlong}
\affiliation{Department of Physics, Zhejiang Normal University, Jinhua 321004, China}
\author{Shu Chen}
\thanks{schen@iphy.ac.cn}
\affiliation{Beijing National
Laboratory for Condensed Matter Physics, Institute of Physics,
Chinese Academy of Sciences, Beijing 100190, China}
\affiliation{School of Physical Sciences, University of Chinese Academy of Sciences, Beijing, 100049, China}
\affiliation{Collaborative Innovation Center of Quantum Matter, Beijing, China}

\begin{abstract}
We study a one-dimensional quasiperiodic system described by the Aubry-Andr\'{e}  model in the small wave vector limit and demonstrate the existence of almost mobility edges and critical regions in the system. It is well known that the eigenstates of the Aubry-Andr\'{e}  model are either extended or localized depending on the strength of incommensurate potential $V$ being less or bigger than a critical value $V_c$, and thus no mobility edge exists. However, it was shown in a recent work that this conclusion does not hold true when the wave vector $\alpha$ of the incommensurate potential is small, and for the system with $V<V_c$, there exist almost mobility edges at the energy $E_{c_{\pm}}$, which separate the robustly delocalized states from ''almost localized" states. We find that, besides $E_{c_{\pm}}$, there exist additionally another energy edges  $E_{c'_{\pm}}$, at which abrupt change of  inverse participation ratio occurs. By using the inverse participation ratio and carrying out multifractal analyses, we identify the existence of critical regions among $|E_{c_{\pm}}| \leq |E| \leq |E_{c'_{\pm}}|$ with the almost mobility edges $E_{c_{\pm}}$ and $E_{c'_{\pm}}$ separating the critical region from the extended and localized regions, respectively. We also study the system with $V>V_c$, for which all eigenstates are localized states, but can be divided into extended, critical and localized states in their dual space by utilizing the self-duality property of the Aubry-Andr\'{e} model.
\end{abstract}
\maketitle
\section{Introduction}
In recent years the localization to delocalization transitions in one-dimensional (1D) quasiperiodic systems have attracted intensive attentions \cite{Modugno-review,Sarma,Qian,Roati,Deissler,Lahini,Roscilde,Orso,Roux,MBL}. This is not only due to the experimental observation of localization transition in the 1D quasiperiodic systems, e.g., ultracold atoms in incommensurate optical lattices \cite{Roati,Deissler} and light waves in quasiperiodic photonic lattices \cite{Lahini}, but also the 1D quasiperiodic systems having already become an important platform for studying interplay effects of controllable disorder and interactions \cite{Roscilde,Orso,Roux,MBL,MBL1,MBL2,Chen1,WDS,hui,Wang,Gao} and topological states in 1D \cite{Lang,Kraus}.  Some interesting examples include the study of quantum many-body localization in quasiperiodic lattices by considering the interaction effect \cite{MBL,MBL1,MBL2} and the topological superconductor to Anderson localization transition in 1D  incommensurate lattices by adding p-wave pairing \cite{Chen1,WDS,hui,Wang}.

As one of the simplest examples of 1D quasiperiodic systems, the Aubry-Andr\'{e} (AA) model \cite{AA} described by
\begin{equation}
\begin{aligned}
&H = t \sum_{i}(\hat{c}^\dagger_{i}\hat{c}_{i+1} +H.c.)+\sum_{i=1}^{L}V\cos(2\pi\alpha i) \hat{n}_i
\label{aa}
\end{aligned}
\end{equation}
has been extensively studied \cite{Suslov,Sokoloff,Soukoulis,Sarma2,Geisel,Machida,classical,Azbel}, where $\hat{c}_{i}$ is a fermionic annihilation operator, $\hat{n}_i=\hat{c}^\dagger_{i}\hat{c}_{i}$ is the particle number operator, $L$ is the size of the system, $t$ is the nearest neighbor hopping amplitude, $V$ is the strength of the incommensurate potential, and $\alpha$ is an irrational number. 
Based on the self-duality property of the model and using the Thouless formula for the localization length \cite{Thouless}, Aubry and Andr\'{e} have shown that all the eigenstates of the model are extended when $V < 2t$, whereas all the eigenstates are localized when $V > 2t$ \cite{AA}. At the self-duality point $V = 2t$, numerical results unveiled that all the eigenstates are critical \cite{kohmoto83,epl}. Although in most studies the irrational number $\alpha$ is chosen to be the golden mean, i.e., $\alpha=\frac{\sqrt{5}-1}{2}$, the argument of Aubry and Andr\'{e} does not depend on the special choice of $\alpha$ and holds true for almost all the  incommensurate systems as long as $\alpha$ is an irrational number. Therefore it is widely believed that all the eigenstates of the AA model with $V < 2t$ are extended for any irrational $\alpha$. However, a recent work by Zhang et. al. \cite{Zhang} numerically verified that there exist two mobility edges at $E_{c_{\pm}}= \pm |2t-V|$ for $V<2t$ when $\alpha$ is a very small irrational number ($\alpha \ll 1 $), which is consistent with the conclusion obtained by using semiclassical analysis \cite{Wilkinson}. When $E_{c_{-}} \leq E \leq E_{c_{+}} $, the eigenstates of the system are extended, whereas they are localized when $E>E_{c_{+}}$ or $E<E_{c_{-}}$. This result seems to be in contradiction with the conclusion of Aubry and Andr\'{e}, which does not support the existence of mobility edges in the AA model. Motivated by these studies \cite{Wilkinson,Zhang}, in this work we revisit the AA model with $\alpha \ll 1$ and scrutinize the properties of wave functions in the whole region of spectrum. Besides the verification of the existence of almost mobility edges $E_{c_{\pm}}$,  out of our expectation,  we find another almost mobility edges $E_{c'_{\pm}}$ and the existence of critical regions at $E_{c'_{-}} \leq E \leq E_{c_{-}} $ and $E_{c_{+}} \leq E \leq E_{c'_{+}} $ for the system with $V < 2t$, where the edges $E_{c_{+}}$ and $E_{c'_{+}}$ separate the critical regions from the extended regions and localized regions, respectively. The critical regions can be distinguished from the localized and extended regions by the different scaling behaviors of wave functions under the finite size analysis. Following the notation in Ref.~\cite{Zhang}, here we name the edges $E_{c_{\pm}}$ as almost mobility edges and, for convenience, also call $E_{c'_{\pm}}$ as almost mobility edges.

This paper is organized as follows: in Sec. \ref{n2}, we study the wave functions of the AA model with $V<2t$ and $\alpha \ll1$ by using the inverse participation ratio (IPR) and find the existence of almost mobility edges $E_{c_{\pm}}$ and $E_{c'_{\pm}}$. The eigenstates of the system are divided into three regions, i.e., the extended, critical and localized region, separated by $E_{c_{\pm}}$ and $E_{c'_{\pm}}$. The three regions can be distinguished by studying the wave function distributions in different regions and performing the multifractal analysis.  In Sec. \ref{n5}, by analyzing the wave functions for the system with $V=2t$ and $V>2t$, we give the phase diagram in the whole regime of $V$. Finally, we give a brief summary in Sec. \ref{n6}.

\section{Almost mobility edges and multifractal analysis of critical wave functions}
\label{n2}
For the AA model described by Eq.~(\ref{aa}), the general eigenstate with eigenvalue $E_n$ is given by $|\Psi_n\rangle=\sum_{i}\psi_{n,i}c_i^{\dagger}|0\rangle$, where $\psi_{n,i}$ is the amplitude of the $nth$ eigenstate at the $ith$ site and $\sum_i|\psi_{n,i}|^2=1$. By using the eigenvalue equation $H|\Psi_n\rangle=E_n|\Psi_n\rangle$, we can obtain the following Harper equation \cite{Harper}
\begin{equation}
\begin{aligned}
t(\psi_{i+1}+\psi_{i-1})+V\cos(2 \pi \alpha i)\psi_{i}=E\psi_{i} .
\label{harper}
\end{aligned}
\end{equation}
Here we leave out the subscript $n$ because the above equation is independent of the specific eigenstate. For convenience, we shall take the hopping amplitude $t=1$ to be the unit of energy and use open boundary conditions unless otherwise stated.
\begin{figure}
\includegraphics[height=80mm,width=80mm]{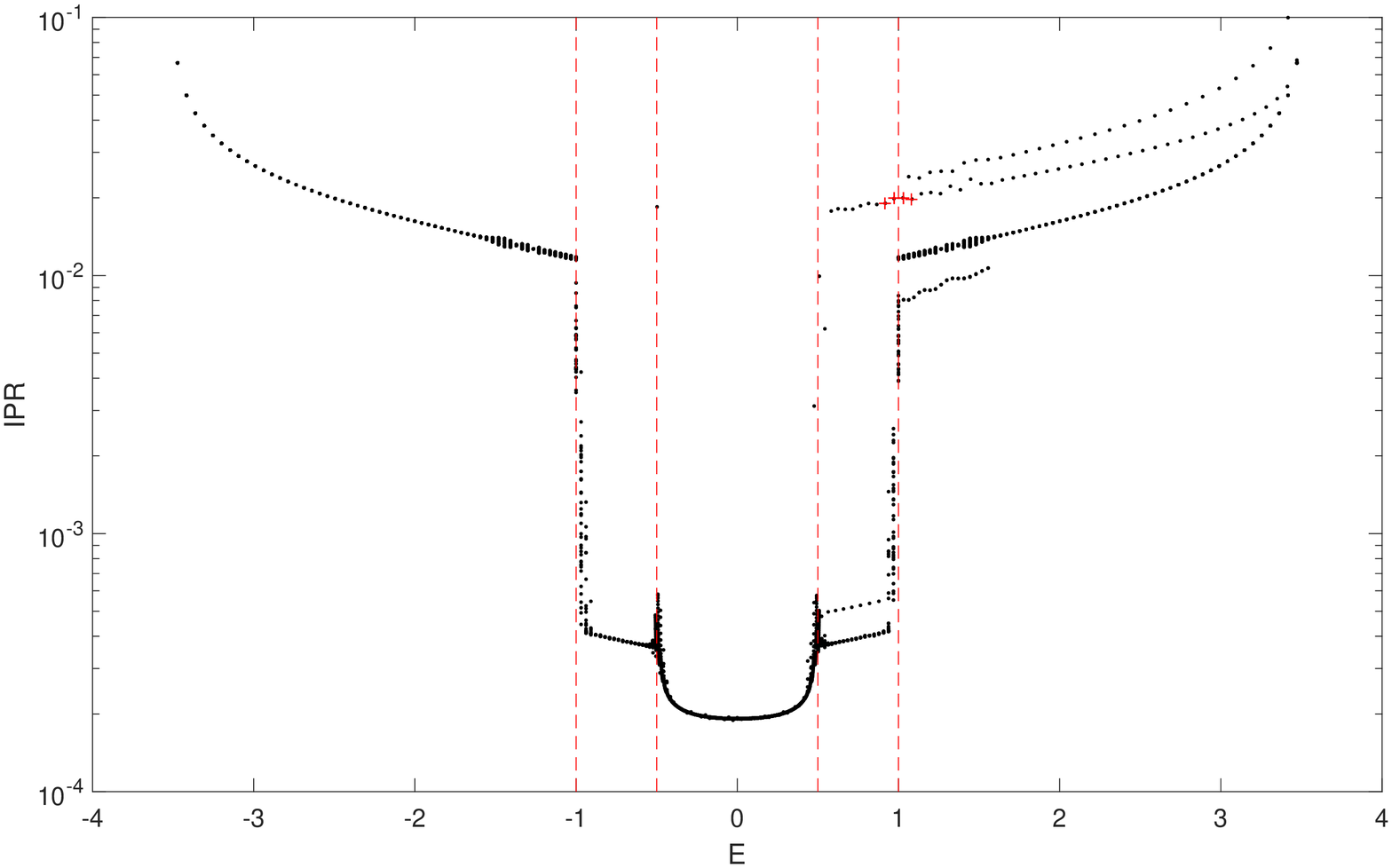}
\caption{\label{IPR}
 (Color Online) The $IPR$ versus $E$ for the system with $V=1.5$, $\alpha=\frac{1}{62\pi}$ and $L=7998$. Here ``Red crosses" represent the IPR of the eigenstates corresponding to states in Fig.~\ref{001} (d), (e), (f) and (g).  "Red dotted lines" correspond $E=-1$, $E=-0.5$, $E=0.5$ and $E=1$.}
\end{figure}

To get an intuitive view of the existence of almost mobility edges in the AA model with small $\alpha$, we consider a concrete system with $V=1.5$, $L=7998$ and $\alpha=\frac{1}{62\pi}$ and calculate the IPR of the system's eigenstate. The IPR \cite{IPR1,IPR2} for the $nth$ eigenstate $|\Psi_n\rangle$ is defined as
\begin{equation}
\begin{aligned}
IPR_n=\sum_{i=1}^L|\psi_{n,i}|^4 .
\label{ipr}
\end{aligned}
\end{equation}
For an extended eigenstate, IPR scales like $L^{-1}$, which tends to $0$ for large $L$, whereas the IPR tend to a finite value $O(1)$ for a localized state. For a critical state, IPR behaves like $L^{-\theta}$, where $0<\theta<1$ depends on the wave-function's multifractal structure. In Fig.~\ref{IPR}, we show IPR as a function of eigenvalues and find sudden changes occurring at $E_{c_{\pm}}=\pm 0.5$,  which is consistent with the prediction of emergence of the almost mobility edges \cite{Zhang} at $E_{c_{\pm}}= \pm |2t-V|$. Moreover, we also observe that obvious sudden changes occur at about $E_{c'_{\pm}}=\pm 1$ in Fig.~\ref{IPR}, which suggests some differences of the wave functions between $|E|>1$ and $0.5<|E|<1$.

\begin{figure}
\includegraphics[height=120mm,width=80mm]{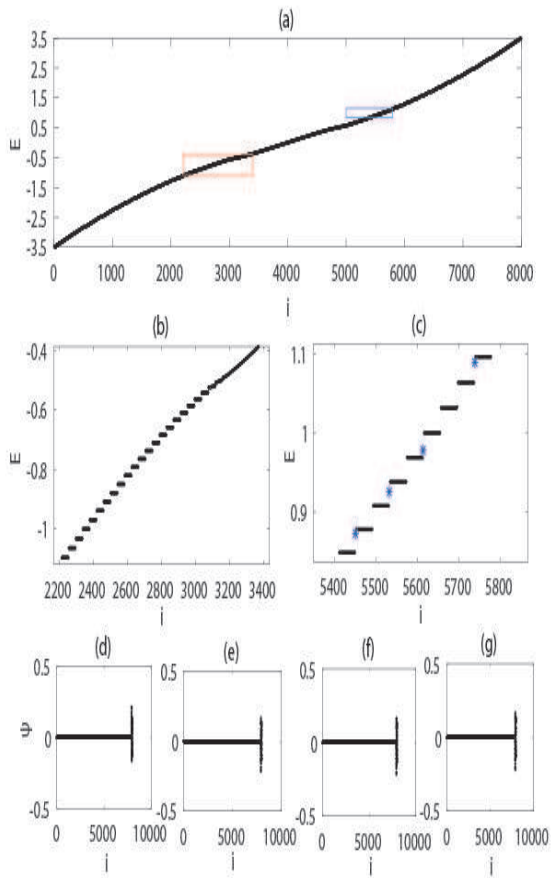}
\caption{\label{001}
 (Color Online)(a) Eigenenergies in ascending order for the system with $V=1.5$, $L=7998$ and $\alpha=\frac{1}{62\pi}$. Enlarged level distributions (b) from $-1.1$ to $-0.4$ (corresponding to the red square in (a)),  and (c) from $0.85$ to $1.1$ (corresponding to the blue square in (a)). Wave function of (d) the $5531th$ eigenstate (corresponding to $E=0.9182$), (e) the $5613th$ eigenstate (corresponding to $E=0.9722$), (f) the $5655th$ eigenstate (corresponding to $E=1.0274$), (g) the $5737th$ eigenstate (corresponding to $E=1.0839$). These states correspond to the states marked by labels $*$ in gaps from bottom to top shown in (c).}
\end{figure}

On the right side of Fig.~\ref{IPR}, we notice that, apart from the main skeleton of the figure, there exist some points, e.g., points marked by red crosses, which have bigger IPRs than those of their neighbor ones. We note that these states corresponds to edge states of the system \cite{Lang,Kraus}.  To make it clear and unveil the properties of spectrum in different regions, we display the spectrum of the system in Fig.~\ref{001}(a). We can find that the energy level distribution for $E>-0.5$ is obviously different from that for $E<-0.5$ (where $-0.5$ corresponding to $E_{c_{-}}=-|2-V|$) by enlarging the left square in Fig.~\ref{001}(a) as shown in Fig.~\ref{001}(b). It is clear that there exist a series of  degenerate states separated by gaps for $E<-0.5$, but no degenerate state exits for $-0.5<E<0.5$. Similar change of energy level distribution is found at $E_{c_{+}}=0.5$, above which the spectrum is composed of degenerate levels separated by gaps. Due to the existence of degenerate levels, the argument of Aubry and Andr\'{e} by using Thouless formula \cite{AA,Thouless} doesn't apply to the case of $\alpha \ll 1$  in the regions of $|E|>0.5$, and thus the conclusion about the absence of mobility edges in the general AA model does not hold true for the case of $\alpha \ll 1$.  Additionally, there appear some edge states as indicated by symbols of ¡°star¡± located at gaps in Fig.~\ref{001}(c), which is the magnification of the right square in Fig.~\ref{001}(a). As shown in Fig.~\ref{001}(d), (e), (f), (g), the four eigenstates corresponding to states marked by stars in Fig.~\ref{001}(c) are localized at the boundary of the system. The IPRs of these edge states correspond to ``red cross" in Fig.~\ref{IPR} and there is a one-to-one correspondence between all isolated points whose eigenstates' IPRs are obviously bigger than their neighbor eigenstates' IPRs and these edge states.

\begin{figure}
\includegraphics[height=150mm,width=80mm]{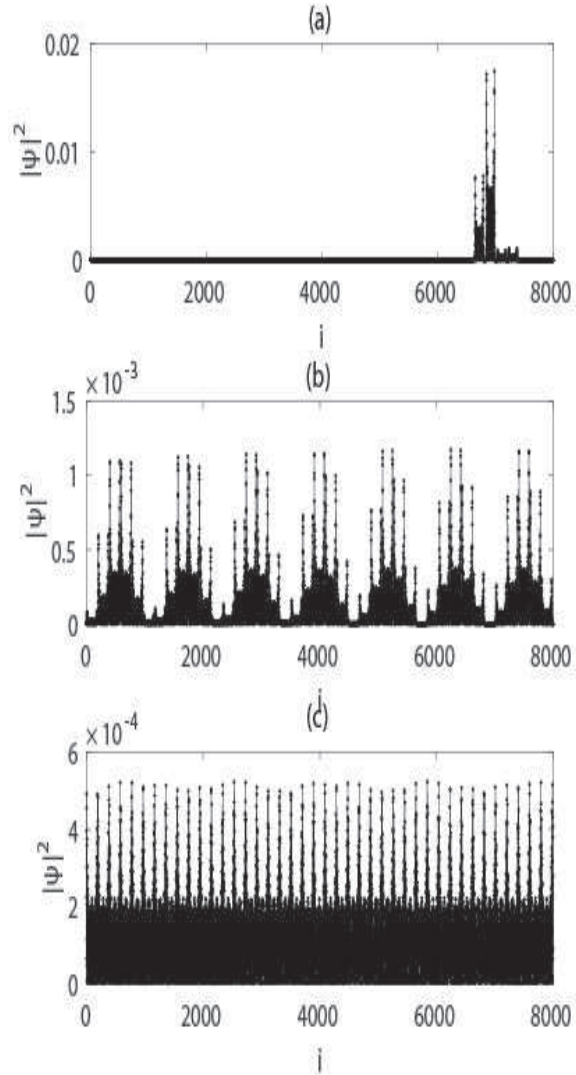}
\caption{\label{003}
The probability distribution of the eigenstate corresponding to (a) $E=-1.0000$ (the $2378th$ eigenvalue), (b) $E=-0.5642$ (the $3000th$ eigenvalue) and (c) $E=-0.3715$ (the $3400th$ eigenvalue) for the system with $ V=1.5$, $L=7998$ and $\alpha=\frac{1}{62\pi}$.}
\end{figure}

To see the difference of distributions of wave functions in different regions, we show the distribution of the $2378th$ eigenstate (corresponding to $E=-1.0000$) in Fig.~\ref{003}(a), the $3000th$ eigenstate (corresponding to $E=-0.5642$) in Fig.~\ref{003}(b) and the $3400th$ eigenstate (corresponding to $E=-0.3715$) in Fig.~\ref{003}(c). While the state in Fig.~\ref{003}(a) is localized, the state in Fig.~\ref{003}(c) is extended. It is shown that the wave function in Fig.~\ref{003}(b) displays different behavior from the localized and extended states, and we shall unveil this state exhibiting multifractal feature by using box-counting method.

\begin{figure}
\includegraphics[height=120mm,width=80mm]{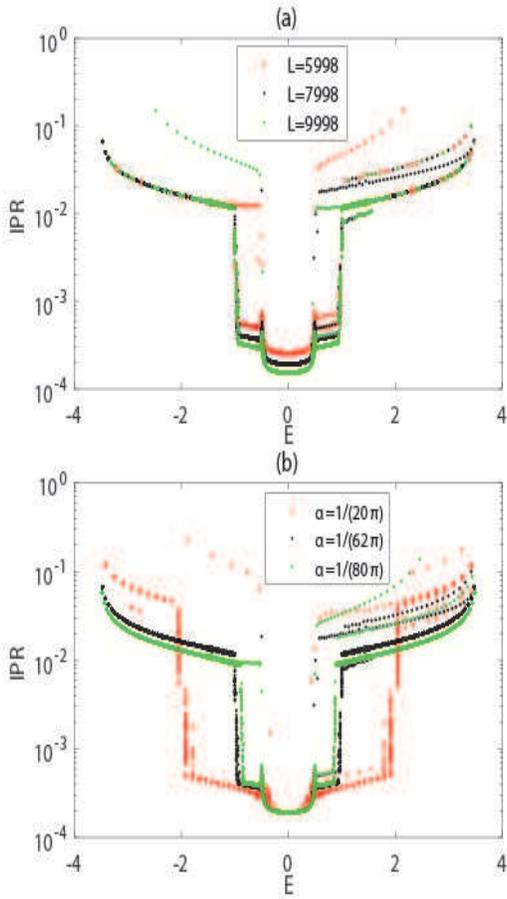}
\caption{\label{002}
 (Color Online) The $IPR$ versus $E$ for the system with $V=1.5$, (a) $\alpha=\frac{1}{62\pi}$ and different lattice size ($L=5998$, $7998$ and $9998$, respectively), (b) $L=7998$ and different $\alpha$, ($\alpha=\frac{1}{20\pi}$, $\frac{1}{62\pi}$ and $\frac{1}{80\pi}$, respectively).}
\end{figure}

\begin{figure}
\includegraphics[height=120mm,width=80mm]{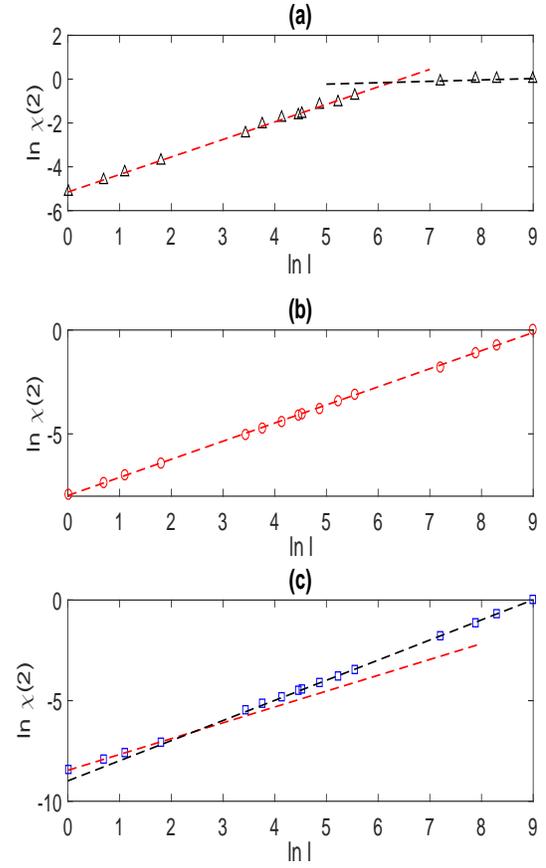}
\caption{\label{004}
 $ln\chi (2)$ as a function of $lnl$ for the system with $V=1.5$, $L=7998$, $\alpha=\frac{1}{62\pi}$ at (a) $E=-1.0000$ (the $2378th$ eigenvalue), (b) $E=-0.5642$ (the $3000th$ eigenvalue) and (c) $E=-0.3715$ (the $3400th$ eigenvalue).}
\end{figure}

In Fig.~\ref{002}(a), we show the IPR as a function of the energy for systems with $V=1.5$, $\alpha=\frac{1}{62\pi}$ and different lattice sizes ($L=5998$, $7998$ and $9998$). We find that sudden changes occur at about $E= \pm 0.5$ and $\pm 1$, corresponding to $E_{c_{\pm}}$ and  $E_{c'_{\pm}}$, respectively, which are independent of the lattice size. There are also some eigenstates whose $IPR$s are obviously bigger than their neighbor eigenstates' $IPR$s, similar to the case as shown in Fig.~\ref{IPR} and Fig.~\ref{001}.
In Fig.~\ref{002}(b), we plot the $IPR$ versus the energy $E$ for systems with $V=1.5$, $L=7998$ and different $\alpha$ ($\alpha=\frac{1}{20\pi}$, $\frac{1}{62\pi}$ and $\frac{1}{80\pi}$). While $E_{c_{\pm}}= \pm 0.5$ do not change with the change of  $\alpha$,   $E_{c'_{\pm}}$ are not universal constants as $|E_{c'_{\pm}}|$ decrease with the decrease of $\alpha$.

\begin{figure}
\includegraphics[height=80mm,width=80mm]{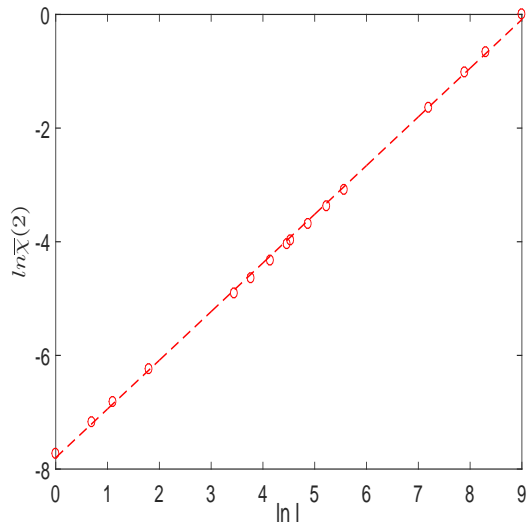}
\caption{\label{005}
 $ln \overline{\chi} (2)$ as a function of $lnl$ for the system with $V=1.5$, $L=7998$ and $\alpha=\frac{1}{62\pi}$ in the region of $-0.9689\leq E\leq -0.5$.}
\end{figure}

Next we study the multifractal properties of the three eigenstates shown in Fig.~\ref{003} by using the box-counting method \cite{epl,Wang,multifractal}. 
Given a wave function which is defined over lattice size $L$ divided into $L/l$ segments with every segment's length $l$, we can define a quantity
\begin{equation}
 \chi_j(q)=\sum_{n=1}^{L/l}[\sum_{i=(n-1)l+1}^{nl}|\psi_{j,i}|^2]^q,
\label{chi}
\end{equation}
where $j$ corresponds to the $jth$ eigenstate.  Multifractal properties can be characterized by a power-law $\chi(q)  \sim  (l/L)^{\tau (q)}$ with the exponent $\tau(q)$ determining the multifractal dimension of this system $D_q = \tau(q)/(q-1)$ \cite{boris,Martin14}. In this work, we set $q=2$  and denote $\chi=\chi(2)$ for abbreviation. In Fig.~\ref{004}, we display the change of  $ln\chi$ as a function of $lnl$ for states shown in Fig.~\ref{003}.
For the state in Fig.~\ref{004}(a), we see that $\chi$ follows a power law $\chi \sim l^{D_2}$ with $D_2=0.7998$ when the length $l$ is smaller than a length, whereas the data for lengths larger than this length is approximated by a line with the slope $D_2\approx 0$, suggesting that the state is localized above the localization length.
For the state in Fig.~\ref{004}(b), $ln\chi$ is a linear function of $lnl$ in the whole lattice size and the slope of the straight line gives the multifractal dimension $D_2=0.8718$. In this case, the wave function shows multifractal feature extending to all length scales of the system, indicating that the state is a critical state.
We note that states in the region $E_{c'_{-}} \leq E \leq E_{c_{-}} $ exhibit similar behaviors as the state in Fig.~\ref{004} (b). By carefully studying all eigenstates by using this box-counting method, we obtain that $E_{c'_{\pm}}=\pm 0.9689$ for $V=1.5, L=7998, \alpha=\frac{1}{62\pi}$, which is actually independent of the lattice size. When the lattice size $L$ increases, the degeneracy also increases, but the eigenenergies of the degenerate eigenstates don't change. For the state in Fig.~\ref{004}(c), we see that $\chi$ follows a power law $\chi \sim l^{D_2}$ with $D_2=0.7867$ when the length $l$ is smaller than a length, whereas the data is approximated by a line with the slope $D_2= 1$ when $l$ is larger than the length, suggesting that the state is extended. Given
the average $\chi$ defined as
\begin{equation}
 \overline{\chi}(2)=\frac{1}{n}\sum_{j=m+1}^{m+n}\chi_j(2),
\label{chi3}
\end{equation}
where $m$ is the number of the eigenvalues corresponding to $E<0.9689$ and $n$ is the number corresponding to $-0.9689\leq E\leq-0.5$,
we display the change of  $ln\overline{\chi}$ as a function of $lnl$ in Fig.~\ref{005}, indicating that $ln \overline{\chi}$ is a linear function of $lnl$ in the whole lattice size for $-0.9689\leq E\leq-0.5$.

\begin{figure}
\includegraphics[height=80mm,width=80mm]{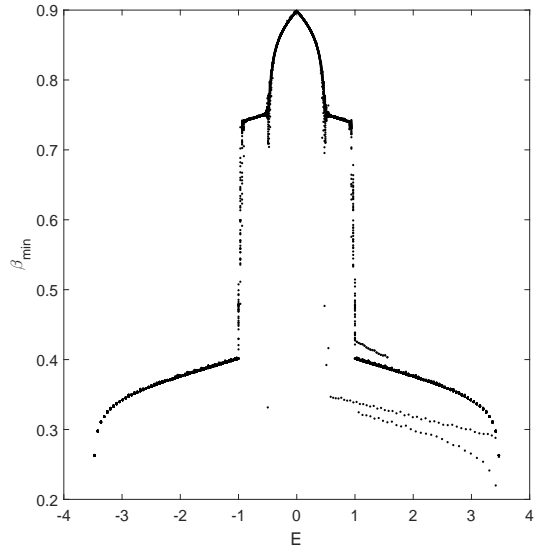}
\caption{\label{006}
 $\beta_{min}$ as a function of $E$ for the system with $V=1.5$, $L=7998$ and $\alpha=\frac{1}{62\pi}$.}
\end{figure}

To further understood the properties of the critical region, we study the scaling behavior of wave functions by using another multifractal analysis \cite{Wang,kohmoto,hui}. We discuss the probability measure at the lattice site $i$ given by $p_{n,i}=|\psi_{n,i}|^2$ ($i=1,2,\cdots,L$) for the $nth$ eigenstate, which is normalized to unity that $\sum_ip_{n,i}=1$. The scaling index $\beta_{n,i}$ for $p_{n,i}$ is defined by
\begin{equation}
 p_{n,i}=L^{-\beta_{n,i}}.
\label{scaling}
\end{equation}
We can use the minimum value of the index $\beta$ to identify the extended, critical or localized wave function, which takes $\beta_{min}=1$ (extended), $0<\beta_{min}<1$ (critical) or $\beta_{min}=0$ (localized) in the thermodynamic limit $L \rightarrow \infty$. In Fig.~\ref{006}, we examine the value of $\beta_{min}$ for all the wave functions of the system with $V=1.5$, $L=7998$ and $\alpha=\frac{1}{62\pi}$. We find three obviously different regions at $|E|>0.9689$, $0.5\leq |E|\leq 0.9689$ and $0<|E|<0.5$, respectively, which is consistent with the Fig.~\ref{IPR}.
\begin{figure}
\includegraphics[height=80mm,width=80mm]{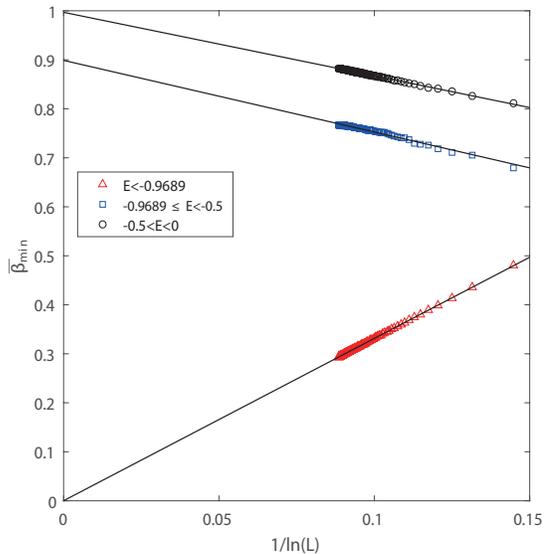}
\caption{\label{007}
 $\overline{\beta}_{min}$ as a function of $1/ln(L)$ for the system with $V=1.5$ and $\alpha=\frac{1}{62\pi}$.}
\end{figure}

To extract  $\beta_{min}$ in the thermodynamic limit, we need study the change of $\beta_{min}$ as a function of lattice size $L$ and make the finite size analysis. As the lattice size is increased, the number of eigenstates increases too. Therefore, we are not able to study the change of $\beta_{min}$ of a given state versus $L$ except for the lowest, middle and the highest states.  To overcome this problem, we consider the mean $\beta_{min}$ of the localized, critical and extended region discussed above as
\begin{eqnarray}
 \overline{\beta}_{min} &=& \frac{1}{m}\sum_{j=1}^{m}\beta_{j,min},
 \label{meanchi1} \\
 \overline{\beta}_{min}&=&\frac{1}{n}\sum_{j=1+m}^{n+m}\beta_{j,min},
 \label{meanchi2} \\
 \overline{\beta}_{min} &=& \frac{1}{k}\sum_{j=1+m+n}^{k+n+m}\beta_{j,min},
\label{meanchi3}
\end{eqnarray}
where $\beta_{j,min}$ corresponds the $\beta_{min}$ of the $jth$ eigenstate after removing the edge states, and $m, n, k$ are the number of the eigenstates corresponding to the localized, critical or extended region, respectively.
We plot $\overline{\beta}_{min}$ as a function of $1/ln(L)$ in Fig.~\ref{007}. We find that $\overline{\beta}_{min}$ extrapolates to $0$ for $E<-0.9689$,  about $0.89$ for $-0.9689\leq E\leq-0.5$ and  $1$ for $-0.5<E<0$, which gives clear evidence for the existence of localized, critical and extended region. The boundary between the localized and critical region is $E_{c'\pm}=\pm 0.9689$ and the boundary between the critical and extended region is $ E_{c \pm}= \pm 0.5$ for the system with $V=1.5$ and $\alpha=\frac{1}{62\pi}$.

\begin{figure}
\includegraphics[height=90mm,width=80mm]{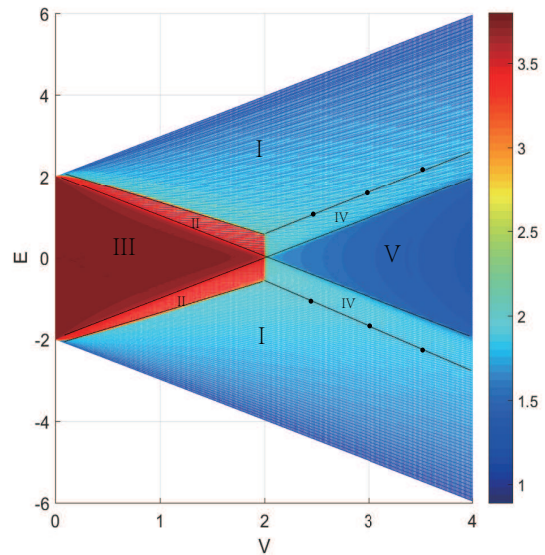}
\caption{\label{008}
 (Color Online) Phase diagram of $E$ versus $V$ for the AA model with $L=7998$ and $\alpha=\frac{1}{62\pi}$. The shading of curves represents the magnitude of the $log_{10}(PR)$ for the corresponding eigenstates. The solid lines of $E_{c_{\pm}}= \pm |2t-V|$ indicate the boundary between the extended region and the critical region and the other solid lines corresponding to $E_{c'_{\pm}}$ indicate the boundary between the critical region and the localized region. The region of $I$, $II$ and $III$ correspond to the localized, critical and extended phase, respectively. States in the region of $IV$ and $V$ are also localized, but the wave functions in their dual space are critical and extended, respectively.}
\end{figure}

\begin{figure}
\includegraphics[height=110mm,width=80mm]{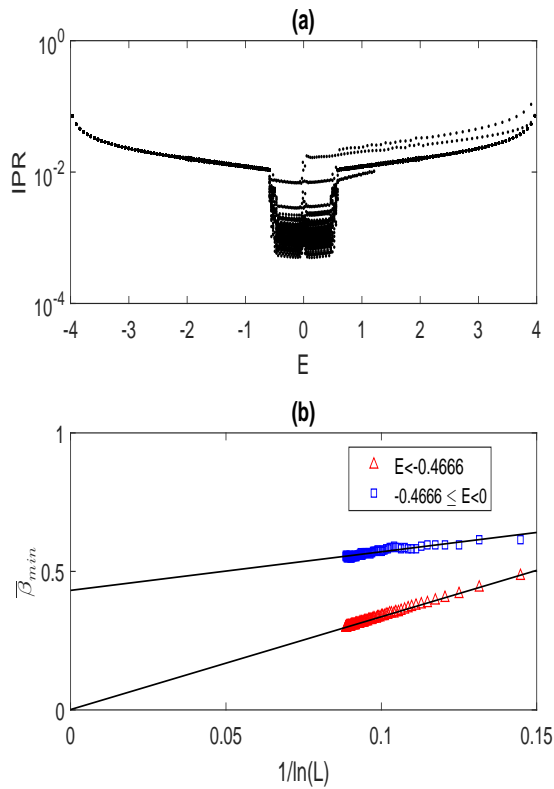}
\caption{\label{009}
 (a) IPR versus $E$, (b) $\overline{\beta}_{min}$ as a function of $1/ln(L)$ for the system with $V=2$, $\alpha=\frac{1}{62\pi}$ and $L=7998$.}
\end{figure}

\begin{figure}
\includegraphics[height=110mm,width=80mm]{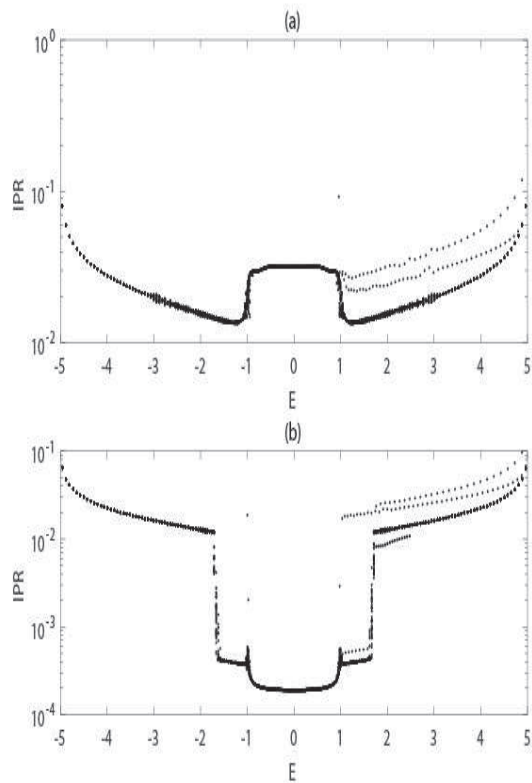}
\caption{\label{010}
 The IPR versus $E$ for the system with $V=3$, $\alpha=\frac{1}{62\pi}$ and $L=7998$ in (a) the real space corresponding to Eq.(\ref{harper}), (b) the dual space corresponding to Eq.(\ref{aa2}).}
\end{figure}

\section{Phase diagram of AA model with $\alpha \ll 1$}
\label{n5}

By applying the above analysis to the system with different $V$, we can determine the phase diagram of the system. For cases with $V<2$, systems display similar behavior as the system with $V=1.5$ discussed in the section II. As shown in Fig.~\ref{008}, the eigenstates of the system with $V<2$ can be divided into three different regions, i.e., region I, II, and III corresponding to the localized, critical and extended phases, respectively. To make it clear, we also display the logarithms of the  participation ratio (PR) of the corresponding eigenstates in the figure, where $PR_{n}=\frac{1}{IPR_n}$. The boundaries between different regions can be detected by the sudden change of $IPR$. Following the same multifractal analysis as in the previous section, we can unveil that states in the region II are critical states.

Now we consider the case of $V=2$ and display the IPR as a function of eigenvalues in Fig.~\ref{009}(a). It is shown that there exists a sudden change near about $E_{\pm}=\pm 0.5$. After studying the wave function distributions and their multifractal properties, we find that there exists two different regions, i.e., the critical and localized region with the boundary at $E_{c'\pm}=\pm 0.4666$, which is consistent with Fig.~\ref{008}. In Fig.~\ref{009}(b), we show $\overline{\beta}_{min}$, defined by Eq.(\ref{meanchi1}) and Eq.(\ref{meanchi2}) after removing the edge states, as a function of $1/ln(L)$. While $\overline{\beta}_{min}$ for $E<-0.4666$ extrapolates to $0$, it extrapolates to  $0.43$ for $-0.4666\leq E<0$.

For the region of $V>2$, it has been shown in Ref. \cite{Zhang} that all eigenstates are localized states. To make it more clear, we display the $IPR$ versus $E$ for the system with $V=3$ in Fig.~\ref{010}(a). Despite the existence of a sudden change of IPR at about $E=\pm 1$, in comparison with the case of $V=1.5$, we find that all values of IPRs are in the order of between $10^{-1}$ and $10^{-2}$, instead of a sharp drop to the order of between $10^{-3}$ and $10^{-4}$ in the band center as shown in Fig.~\ref{IPR}. By using the finite size analysis, we can verify that all eigenstates are localized.

Although the region of $V>2$ corresponds to a localized region without mobility edges \cite{Zhang}, we can still divide it into three regions according to the IPR of wave function in the dual space of the model due to the existence of self duality of the AA model. We can obtain the dual model of Eq.(\ref{harper}) by introducing a transformation
\begin{equation}
 f_m=\frac{1}{\sqrt{L}}\sum_n\psi_n e^{i2\pi \alpha nm}.
\label{transformation}
\end{equation}
It is easy to find that $f_m$ satisfies:
\begin{equation}
\begin{aligned}
\frac{V}{2}(f_{m+1}+f_{m-1})+2t\cos(2 \pi \alpha m)f_m =Ef_m,
\label{aa2}
\end{aligned}
\end{equation}
which has the same form as Eq.(\ref{harper}) except of exchanging the role of $V$ and $t$.
The IPR of wave function in the dual space is then defined by
\begin{equation}
\begin{aligned}
IPR =\sum_{m=1}^L|f_{m}|^4 .
\label{iprd}
\end{aligned}
\end{equation}
In Fig.~\ref{010}(b), we display the IPR in the dual space versus $E$ for the system with $V=3$, which exhibits a similar structure as shown in Fig.~\ref{IPR} for the system with $V<2t$. So even all eigenstates are localized in the original space for the system with $V>2t$, there still exist some hidden differences which can be distinguished in the dual space. As shown in Fig.~\ref{008}, the region of V, IV, and I correspond to extended, critical and localized phase in the dual space, respectively.
For the phase diagram shown in Fig.~\ref{008}, it is interesting to indicate that all of the eigenstates are localized in both the original space and  dual space in the region $I$, are critical in the original space but localized in the dual space in the region $II$, are extended in the original space but localized in the dual space in the region $III$, are localized in the original space but critical in the dual space in the region $IV$, are localized in the original space but extended in the dual space in the region $V$.

\section{Summary}
\label{n6}
In summary, we have studied the 1D incommensurate system described by the AA model with $\alpha\ll 1$ and demonstrate that the system displays quite different behavior from the general AA model with large wave vector $\alpha\sim O(1)$. While all eigenstates of the general AA model with $V<2t$ are extended, we find the existence of almost mobility edges $E_{c_{\pm}}$ and  $E_{c'_{\pm}}$ for the AA model with very small wave vector $\alpha\ll 1$, and the eigenstates are divided into three different regions by these  almost mobility edges. By carrying out multifractal analysis, we identify that the regions among $|E_{c_{\pm}}| \leq |E| \leq |E_{c'_{\pm}}|$  are critical regions with wave functions in these regions displaying self-similar behaviors, whereas the region of $E_{c_{-}} < E  < E_{c_{+}}$ and  of $ E < E_{c'_{-}}$ and $ E > E_{c'_{+}}$ correspond to the extended and localized regions, respectively. We also present the phase diagram of $E$ versus $V$ in the whole region of $V$. Particularly, for the system with $V>2t$, while all eigenstates are localized states, we can divide the states into different regions according to the extended, critical and localized properties in their dual space by utilizing the self-duality property of the AA model.

\begin{acknowledgments}
 The work is supported by NSFC under Grants No. 11425419, No. 11374354 and No. 11174360, and the Strategic Priority Research Program (B) of the Chinese Academy of Sciences  (No. XDB07020000). Gao Xianlong was supported by the NSF of China (Grant No. 11374266) and the Program for New Century Excellent Talents in University.
\end{acknowledgments}

\end{document}